\documentclass[fleqn,10pt]{wlscirep}
\usepackage[utf8]{inputenc}
\usepackage[T1]{fontenc}
\usepackage[caption=false]{subfig} 
\usepackage[normalem]{ulem} 
\usepackage{multirow}
\usepackage[table,xcdraw]{xcolor} 
\usepackage[normalem]{ulem}
\usepackage[absolute,overlay]{textpos}
\usepackage{hyperref}
\usepackage{xr-hyper}
\usepackage{kotex} 
\useunder{\uline}{\ul}{}

\title{Mutual information maximizing quantum generative adversarial networks}

\author[1,2]{Mingyu Lee}
\author[2,3]{Myeongjin Shin}
\author[2,4,*]{Junseo Lee}
\author[2,5,6,$\dagger$]{Kabgyun Jeong}

\affil[1]{Department of Computer Science and Engineering, Seoul National University, Seoul 08826, Korea}
\affil[2]{Team QST, Seoul National University, Seoul 08826, Korea}
\affil[3]{School of Computing, KAIST, Daejeon 34141, Korea}
\affil[4]{Quantum AI Team, Norma Inc., Seoul 04799, Korea}
\affil[5]{Research Institute of Mathematics, Seoul National University, Seoul 08826, Korea}
\affil[6]{School of Computational Sciences, Korea Institute for Advanced Study, Seoul 02455, Korea}

\affil[*]{Email: harris.junseo@gmail.com}
\affil[$\dagger$]{Email: kgjeong6@snu.ac.kr}


\begin{abstract}
One of the most promising applications in the era of Noisy Intermediate-Scale Quantum (NISQ) computing is quantum generative adversarial networks (QGANs), which offer significant quantum advantages over classical machine learning in various domains. However, QGANs suffer from mode collapse and lack explicit control over the features of generated outputs. To overcome these limitations, we propose InfoQGAN, a novel quantum-classical hybrid generative adversarial network that integrates the principles of InfoGAN with a QGAN architecture. Our approach employs a variational quantum circuit for data generation, a classical discriminator, and a Mutual Information Neural Estimator (MINE) to explicitly optimize the mutual information between latent codes and generated samples. Numerical simulations on synthetic 2D distributions and Iris dataset augmentation demonstrate that InfoQGAN effectively mitigates mode collapse while achieving robust feature disentanglement in the quantum generator. By leveraging these advantages, InfoQGAN not only enhances training stability but also improves data augmentation performance through controlled feature generation. These results highlight the potential of InfoQGAN as a foundational approach for advancing quantum generative modeling in the NISQ era.

\end{abstract}
\begin{document}

\flushbottom
\maketitle
\thispagestyle{empty}

\section*{Introduction}
The advancement of classical neural networks has profoundly impacted various domains by enabling sophisticated pattern recognition and data modeling capabilities. Among these developments, Generative Adversarial Networks (GANs)~\cite{goodfellow2014generative} have emerged as a cornerstone of modern generative modeling. GANs have been successfully applied across diverse fields, including high-fidelity image synthesis~\cite{karras2019style}, data augmentation~\cite{antoniou2017data}, and numerous other applications. The training process of GANs is formulated as a minimax optimization problem between the generator and the discriminator, expressed as follows:  
\begin{align}\label{ganeq}  
    \min_G\max_D V(D, G) = \underset{{x\sim p_\text{data}(x)}}{\mathbb{E}}[\log D(x)] + \underset{{z\sim p_z(z)}}{\mathbb{E}}[\log(1-D(G(z)))],  
    \end{align}  
where $ z = (z_1, z_2, \dots, z_n) $ is the latent noise vector sampled from a prior distribution $ p_z(z) $, $ D(x) $ denotes the discriminator's confidence that $ x $ is a real sample, and $ D(G(z)) $ represents its assessment of the authenticity of the generated data.  

Despite their remarkable success, GANs suffer from inherent limitations. One major issue is \textbf{mode collapse}, where the generator fails to capture the full diversity of the data distribution and produces a restricted set of outputs~\cite{che2016mode, dumoulin2016adversarially, salimans2016improved, saatci2017bayesian, nguyen2017dual, lin2018pacgan, ghosh2018multi}. This is because the objective function of the generator only considers whether it can fool the discriminator, without ensuring diversity in the outputs.
Another critical limitation is the lack of \textbf{feature disentanglement}, where variations in the latent code fail to produce distinct and interpretable changes in the generated samples. This makes it difficult to control specific attributes of the output, limiting the model’s applicability in tasks that require semantic manipulation or interpretability.
To mitigate these issues, various enhancements have been proposed. Among them, Information Maximizing Generative Adversarial Networks (InfoGAN)~\cite{chen2016infogan} extend the conventional GAN framework by introducing a latent code that explicitly influences the generator's output. InfoGAN partitions the input noise vector $ z $ into two components: a standard noise component and a structured latent code $c$, expressed as $ z = (z, c) = (z_1, \dots, z_{n-m}, c_1, \dots, c_{m}) $, where $ m $ denotes the dimension of the latent code space. The key objective is to maximize the mutual information $ I(c; G(z,c)) $, ensuring both high diversity in the generated outputs and that the latent code effectively controls distinct attributes of the samples. This is achieved by modifying the standard GAN minimax loss function as follows:
\begin{equation}\label{infogan}  
    \min_G\max_D V_I(D, G)=V(D, G) - \beta\cdot I(c; G(z,c)).  
\end{equation}  

The coefficient $ \beta > 0 $ regulates the contribution of mutual information in the training process. In InfoGAN, the architecture of the discriminator was modified to approximate $I(c; G(z,c))$. However, a more efficient method for estimating MI, known as the Mutual Information Neural Estimator (MINE)~\cite{belghazi2018mine}, was later introduced. MINE is based on the principles of Kullback-Leibler (KL) divergence~\cite{kullback1997information} and the Donsker-Varadhan (DV) lower bound~\cite{von2013mathematische}. By deriving the estimation formula using the KL divergence, MINE demonstrated that MI can be effectively estimated using a simple neural network architecture.


This unsupervised learning paradigm between the generator's output and the latent code leads to striking results. InfoGAN not only mitigates mode collapse but also enables fine-grained control over generated features by independently varying the latent code. This is because it forces changes in latent code to be meaningfully reflected in the generated output. In contrast, traditional GANs merely attempt to map input noise to the target distribution, often resulting in highly entangled representations. Consider the example of generating digits $ 0 $–$ 9 $ from the MNIST dataset. While a standard GAN takes input noise and produces random digits, InfoGAN allows explicit control over the generated digit, as well as attributes such as stroke thickness and tilt, by adjusting the latent code. The effectiveness of simple unsupervised learning based on mutual information maximization, combined with minimal modifications to the GAN architecture, has sparked significant interest in InfoGAN and its potential applications.

With the rapid advancement of quantum computing, there has been a growing surge of interest in Quantum Machine Learning (QML)~\cite{biamonte2017quantum, schuld2015introduction}. As a promising direction within this field, Quantum Generative Adversarial Networks (QGANs) have been introduced~\cite{dallaire2018quantum, lloyd2018quantum} to leverage quantum circuits for data generation, particularly in high-dimensional spaces, where classical methods often struggle.  

\begin{figure}[t]
   \centering
   \includegraphics[width=10cm]{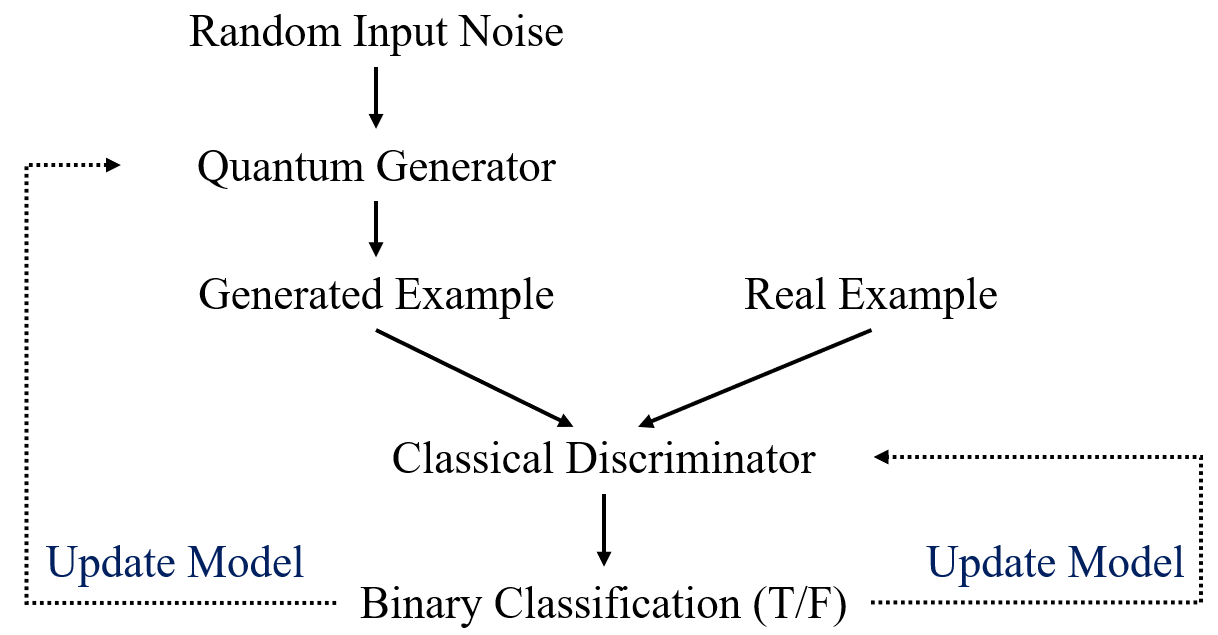}
   \caption{The basic structure of a quantum generative adversarial networks (QGAN).}
   \label{fig:qganst}
\end{figure}

A standard QGAN architecture typically consists of a quantum generator and a classical discriminator, as depicted in Figure~\ref{fig:qganst}. While fully quantum implementations, in which both components are quantum, are theoretically possible, the hybrid quantum-classical approach remains more practical in the NISQ (Noisy Intermediate-Scale Quantum) era due to the technical constraints of current quantum hardware. By combining quantum data generation with classical evaluation, this hybrid strategy enables efficient training while mitigating the limitations of near-term quantum devices. The learning process in QGAN follows the same fundamental principles as classical GANs, with the primary distinction being the replacement of the classical generator with a quantum generator, typically implemented using a variational quantum circuit (VQC).


QGANs have a wide range of potential applications, including modeling general probability distributions~\cite{zoufal2019quantum}, quantum state preparation~\cite{kim2024hamiltonian}, drug discovery~\cite{li2021quantum}, and image processing~\cite{xiao2023practical}, among others. These examples highlight QGANs as a promising tool for both classical and quantum data domains. However, in conventional frameworks, the output of QGANs is solely determined by the input noise, which limits the ability to explicitly control or manipulate desired attributes of the generated data. In particular, the quantum processes involved in QGANs are more difficult to investigate than those in classical neural networks. This makes it challenging to track the internal behavior of the model and further complicates the task of capturing the relationship between input and output.

Also, as a direct quantum counterpart of GANs, QGANs inherit well-known challenges such as mode collapse~\cite{dallaire2018quantum, lloyd2018quantum}. As a result, oscillations among a limited set of modes can severely impact training stability, making convergence more difficult~\cite{niu2022entangling}.

Motivated by the success of InfoGAN in addressing the limitations of classical GANs, we aim to investigate whether a similar approach can be applied to QGANs. In this context, this study incorporates the InfoGAN framework into QGANs, introducing a novel InfoQGAN architecture. This approach not only mitigates mode collapse but also enables feature disentanglement within the quantum generator—even at the level of quantum operations—an ability that was previously unattainable in quantum generative models. Through this enhancement, we demonstrate that InfoQGAN significantly improves the expressiveness and stability of quantum generative learning.

\section*{Integration of QGAN and MINE}
We propose a novel quantum machine learning framework that seamlessly integrates QGAN with MINE, leveraging the strengths of both methodologies. The quantum generator follows a standard QGAN structure, with the parameterized quantum circuit (PQC) ansatz constructed using rotation-$Y$ (RY) and controlled-$X$ (CNOT) gates. We use RY and CNOT gates to construct the ansatz because both the input and output data are real-valued ~\cite{zoufal2019quantum}.

\begin{figure}[t]
   \centering
   \includegraphics[width=10cm]{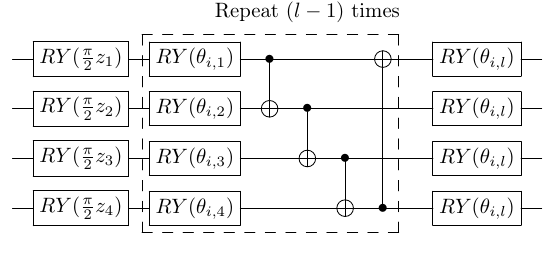}
   \caption{Generator ansatz for a 4-qubit quantum circuit with $l$ layers.}
   \label{fig:ansatz}
\end{figure}

The generation process begins with an input noise vector $ z $, which undergoes an initialization step where an RY gate is applied to each qubit. The rotation angle is set to $ {\pi z}/{2} $, effectively scaling the input noise to a more suitable range. Following this, the generator's trainable parameters $ \theta $ are introduced through a structured ansatz.   For a single-layer ansatz, an RY is first applied to each qubit, followed by cyclic entanglement using CNOT gates. Specifically, CNOT gates are sequentially applied between adjacent qubits, forming a closed ring topology: qubit 0 is entangled with qubit 1, qubit 1 with qubit 2, continuing until qubit $ n-1 $ is connected back to qubit 0. This cyclic entanglement structure, which is depicted in Figure~\ref{fig:ansatz}, ensures strong quantum correlations across the circuit while maintaining an efficient and scalable design. Of course, the ansatz can be modified into different forms while appropriately considering the expressiveness and scale of the circuit, and techniques such as genetic algorithms or quantum architecture search can be introduced when necessary.

In the final state, we measure the expectation values $ \{\langle O_j\rangle\} $ of local observables. Using these values, we compute the model’s $ N $-dimensional output $ G(z) $. To ensure that activation functions remain compatible with quantum computation, they must support gradient evaluation via the parameter shift rule~\cite{mitarai2018quantum, wierichs2022general}. While this study employs simulated quantum circuits, the framework can also be applied to real quantum hardware by leveraging hybrid training methods based on the chain rule. 

Following the InfoGAN framework, the input noise vector $ z $ is decomposed into $ (z, c) $ to incorporate latent code disentanglement. The discriminator and MINE model are implemented as classical feedforward neural networks, both trained via classical backpropagation. The discriminator consists of three hidden layers, each with 50 units, where each layer applies a linear transformation followed by a LeakyReLU activation. A final linear layer with a Sigmoid activation produces a single scalar output for classification.  

The discriminator consists of a fully connected feedforward network composed of four linear layers. The input is first projected to a 50-dimensional hidden space, followed by a LeakyReLU activation. This hidden representation is further transformed through two additional hidden layers, each of size 50, with LeakyReLU activations applied after each transformation. Finally, a linear layer with a sigmoid activation maps the output to a single scalar value, representing the discriminator’s confidence that the input is a real sample.

The MINE architecture follows the original design~\cite{belghazi2018mine}, consisting of two parallel linear layers that separately process the latent code and real/fake data. Each input is mapped to a 50-dimensional space, and their outputs are summed and passed through a ReLU activation. A final linear layer reduces this representation to a single scalar output, which serves as the mutual information estimate.

By replacing InfoGAN's mutual information estimation with MINE, we introduce a novel quantum-classical hybrid model that we refer to as \textbf{InfoQGAN}—a mutual information maximizing QGAN.

\section*{Numerical experiments and results}
We conducted two numerical experiments to compare InfoQGAN with conventional QGAN, as well as their classical counterparts, InfoGAN and GAN. To ensure a fair comparison, the generators in InfoQGAN and QGAN share the same architecture and number of parameters. Similarly, the generators in InfoGAN and GAN are identical. The quantum and classical components also have comparable numbers of parameters. Given the limited number of parameters, we employed a classical generator with the simplest fully connected feedforward neural network. This classical generator consists of three linear layers: the first layer projects the input dimension to a hidden dimension, followed by a sigmoid activation; the second layer maintains the hidden dimension with another sigmoid activation; and the final layer maps the hidden dimension to the output dimension, with a sigmoid activation applied at the end.

To evaluate the effectiveness of InfoQGAN, we designed two tasks. First, we tasked the model with generating a predefined 2D distribution to enable clear visual assessment. Second, we applied the model to augment the Iris dataset~\cite{fisher1936use}, demonstrating its effectiveness in real-world applications. Our study specifically investigates the following key questions:  
\begin{enumerate}  
    \item Is InfoQGAN less susceptible to mode collapse compared to QGAN?  
    \item Can InfoQGAN successfully disentangle generated features, akin to the capabilities observed in InfoGAN?
    \item Can InfoQGAN be effectively utilized in real-world applications such as data augmentation?  
    \item How do the computational and parameter requirements of InfoQGAN compare to those of conventional QGAN?  
\end{enumerate}

The answer to the final question is that InfoQGAN does not require a significantly greater computational overhead compared to the conventional QGAN. A detailed comparison is provided in Tables~\ref{sup-tab:disc-2d-flops},~\ref{sup-tab:mine-2d-flops},~\ref{sup-tab:disc-iris-flops}, and~\ref{sup-tab:mine-iris-flops}, which are included in the supplementary material. Although the overheads depend on the numerical setup, InfoQGAN results in approximately a 6\%--8\% increase in FLOPs and a 7\%--8\% increase in the number of trainable parameters. Moreover, since these requirements pertain solely to the classical part of the architecture, we believe they will not pose a significant burden in the near future, when the quantum part is expected to be the primary bottleneck.

\subsection*{2-dimensional data generation}
All training was conducted on an ideal simulator, and for InfoQGAN and QGAN, we also compared results obtained using the Qiskit GenericBackendV2 simulator (hereafter referred to as the noisy environment). The Qiskit GenericBackendV2 simulator is constructed by randomly sampling gate error rates and T1/T2 times from historical IBM backend data.

The predefined target 2D distribution is a uniform, diamond-shaped region within the $[0,1]$ domain, centered at $(0.6, 0.6)$ with a side length of 0.4.

We used a quantum generator with 5 qubits and 10 layers, resulting in 50 trainable parameters. The generator circuit structure is identical for both InfoQGAN and QGAN. Since the target coordinate range is $[0,1]^2$, the probability of the first two qubits being in the $|1\rangle$ state is directly interpreted as the generated point. InfoGAN and GAN adopt the same input structure as InfoQGAN and QGAN. By setting the hidden dimension to 4, each model contains a total of 54 parameters.  

The generator's input is sampled from $z = (z_1, z_2, z_3, c_1, c_2) \sim \mathcal{U}([-0.5, 0.5])^{5}$, where $\mathcal{U}(S)$ denotes the uniform distribution over the set $S$. As indicated by the notation, we incorporated a two-dimensional latent code space, though this has no effect in the case of QGAN and GAN.  

Given the differing learning speeds of variational quantum circuits and classical neural networks, we set distinct learning rates for each case. For the quantum models, the learning rates for both the generator and MINE were set to 0.001. For the classical models, the generator and MINE were assigned a learning rate of 0.002. The discriminator learning rate was consistently set to 0.0005. The training parameter $\beta$ in Equation~\ref{ganeq} was fixed at 0.2. A summary of the training hyperparameters is provided in Table ~\ref{sup-tab:2d-gen-hparams} in the supplementary material.

To assess the similarity between the generated point distributions, we employed the two-dimensional two-sample Kolmogorov-Smirnov (KS) test~\cite{peacock1983two}. The null hypothesis states that both datasets are drawn from the same distribution, while the alternative hypothesis suggests they originate from different distributions.

An epoch-wise comparison of the generated outputs is shown in Figure~\ref{sup-fig:compare_mode_collapse}, which is included in the supplementary material. In the outputs generated by QGAN across different epochs, samples tend to cluster in specific regions, suggesting that the model captures only a subset of the target distribution. Furthermore, the model frequently oscillates between these modes, leading to reduced training stability. In contrast, InfoQGAN consistently generates outputs that are uniformly distributed across the entire target space throughout training. This indicates that InfoQGAN effectively mitigates mode collapse, whereas QGAN does not.

\begin{figure}[t]
     \centering
     \subfloat[GAN]{
         \includegraphics[width=0.16\textwidth]{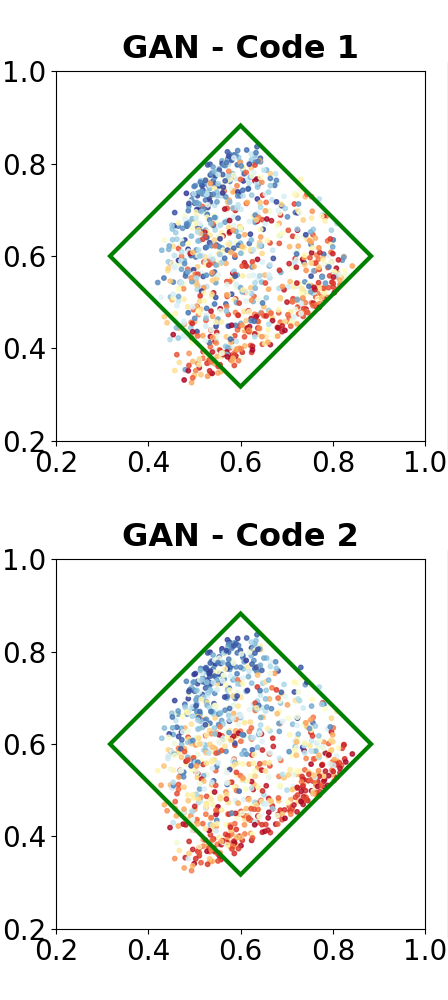}
         \label{fig:disentanglement_GAN}
     }\quad
     \subfloat[InfoGAN]{
         \includegraphics[width=0.16\textwidth]{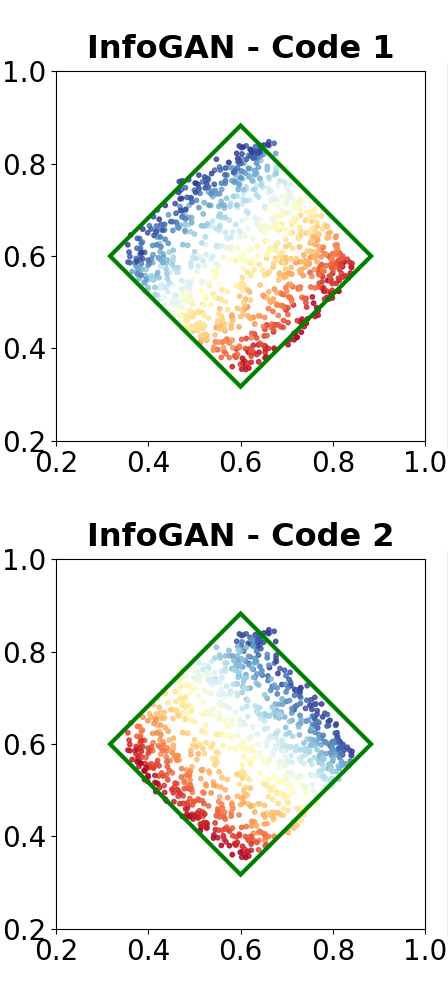}
         \label{fig:disentanglement_InfoGAN}
     }\quad
     \subfloat[QGAN]{
         \includegraphics[width=0.16\textwidth]{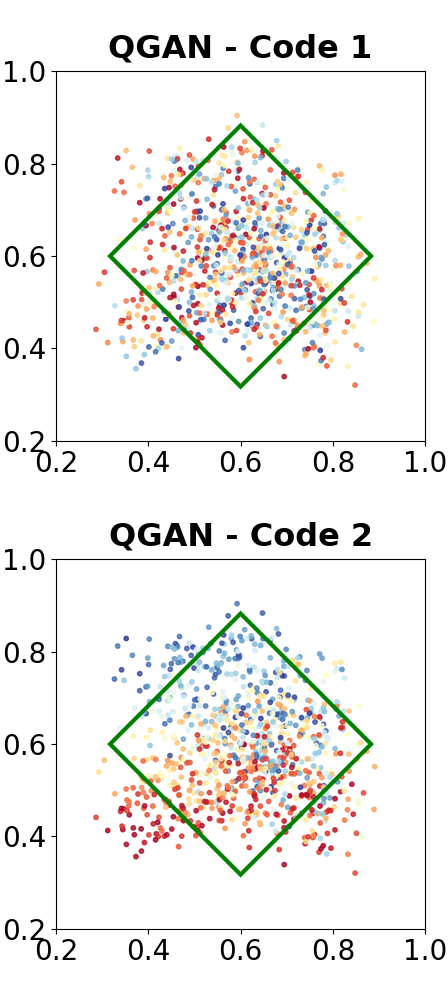}
         \label{fig:disentanglement_QGAN}
     }\quad
     \subfloat[InfoQGAN~(ideal)]{
         \includegraphics[width=0.16\textwidth]{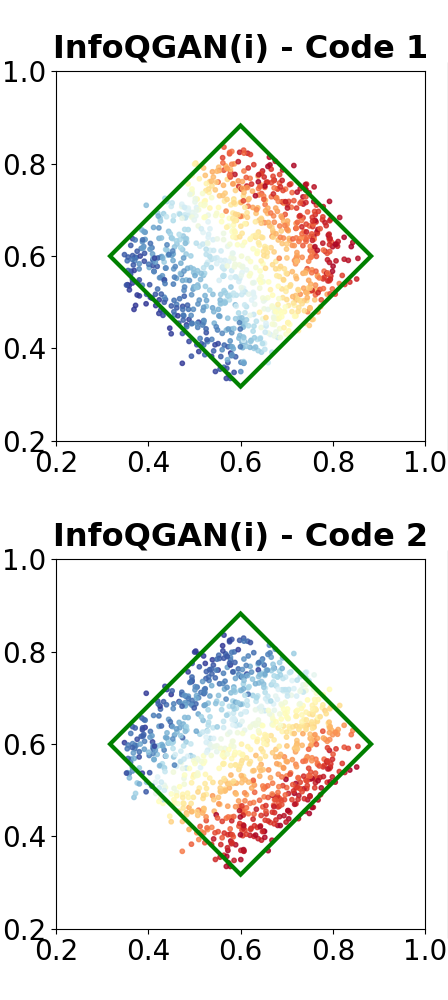}
         \label{fig:disentanglement_InfoQGAN_ideal}
     }\quad
     \subfloat[InfoQGAN~(noisy)]{
         \includegraphics[width=0.2142\textwidth]{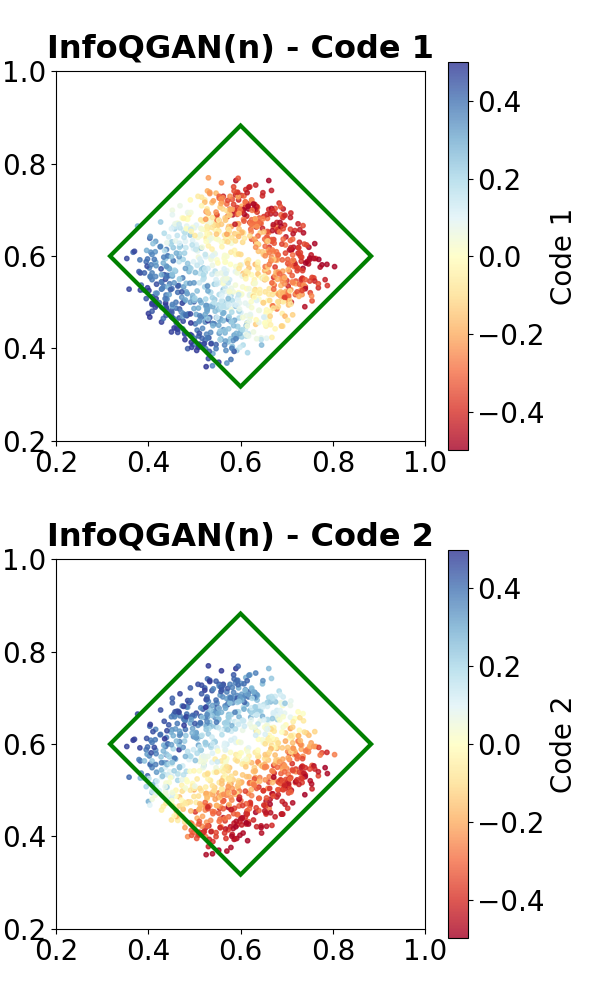}
         \label{fig:disentanglement_InfoQGAN_noisy}
     }
\caption{\textbf{Comparison of latent code disentanglement.} Samples were generated from the epoch with the highest $p$-value during training. InfoQGAN and InfoGAN exhibit superior disentanglement capabilities, as evidenced by the clear color separation of latent codes in the generated samples. InfoQGAN also demonstrates feature disentanglement in a noisy environment; However, due to the noise-induced variance reduction, it produces a smaller-shaped distribution.}
    \label{fig:disentanglement_compare}
\end{figure}

Furthermore, we highlight the remarkable capabilities of InfoQGAN. Figures~\ref{fig:disentanglement_compare} depict the model’s output points, colored according to the input latent code. The distinct color separation observed in InfoQGAN indicates successful feature disentanglement in generating the 2D distribution—an outcome that QGAN fails to achieve. Compared to the classical models, InfoGAN successfully disentangles features, whereas GAN does not. This observation suggests that InfoQGAN effectively disentangles features along distinct dimensions in Hilbert space.

Also, in noisy environments, clear color separation was still observed. However, we would like to clarify why the diamond shape in Figure~\ref{fig:disentanglement_InfoQGAN_noisy} appears smaller than that in the ideal case Figure~\ref{fig:disentanglement_InfoQGAN_ideal}. This phenomenon occurred in both QGAN and InfoQGAN. Due to noise-induced errors, the probability values tended to cluster around 0.5 ~\cite{wang2021noise}, causing InfoQGAN to generate more compact diamond-shaped distributions. Consequently, this led to a significant reduction in the $p$-value, which serves as a statistical test for the equality of distributions, as will be shown later.

We further observed that color separation consistently occurs perpendicularly between the two latent codes. To quantify this, we computed the correlation vectors between the generated $x$ and $y$ coordinates for each code, denoted as $v_{c_1}$ and $v_{c_2}$. We then calculated the angle between these vectors. Additionally, we measured the norm of each vector, $|v_{c_1}|$ and $|v_{c_2}|$, as it reflects the explanatory power of the latent codes and serves as further evidence of feature disentanglement.

A summary of the numerical results is presented in Table~\ref{tab:exp1result}. InfoQGAN achieved a lower Kolmogorov-Smirnov (KS) value and a higher $p$-value compared to QGAN, highlighting its significant advantage in mitigating mode collapse. The angle between $v_{c_1}$ and $v_{c_2}$ was close to $\pi/2$ in InfoQGAN and InfoGAN, unlike in the other models. Additionally, the norm of each correlation vector was higher in InfoQGAN and InfoGAN, further supporting their superior disentanglement capabilities. For InfoQGAN, even in noisy simulations, the separation angles remained sufficiently large, and the explanatory power of each latent code remained high, indicating that the disentanglement ability was largely preserved.

\begin{figure}[t]
     \centering
     \subfloat[Target distribution]{
         \includegraphics[width=0.48\textwidth]{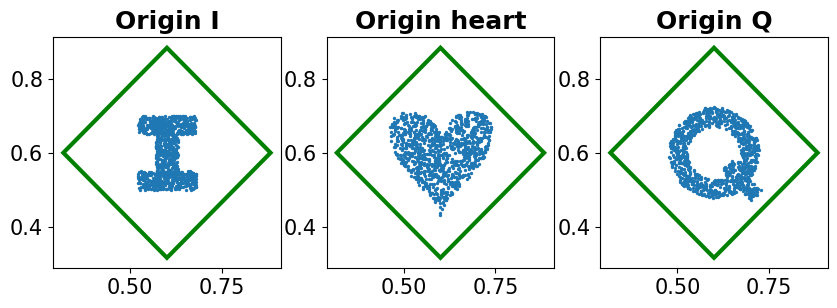}
         \label{fig:2d_custom_origin}
     }\quad
     \subfloat[Transforming from target distribution to code space (InfoQGAN)]{
         \includegraphics[width=0.48\textwidth]{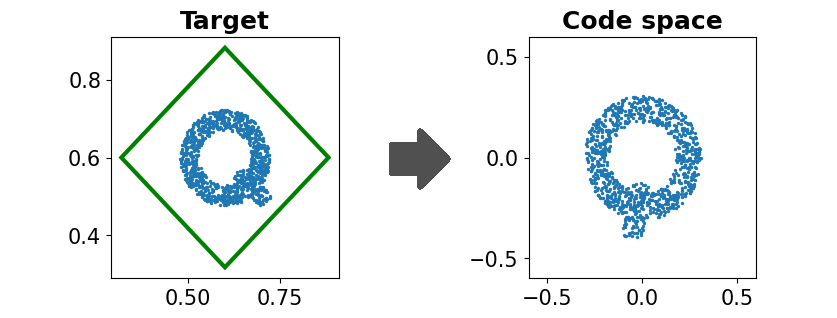}
         \label{fig:2d_custom_InfoQGAN_Q}
     }
    \vspace{0.001cm}
     \subfloat[Distribution generated by QGAN]{
         \includegraphics[width=0.48\textwidth]{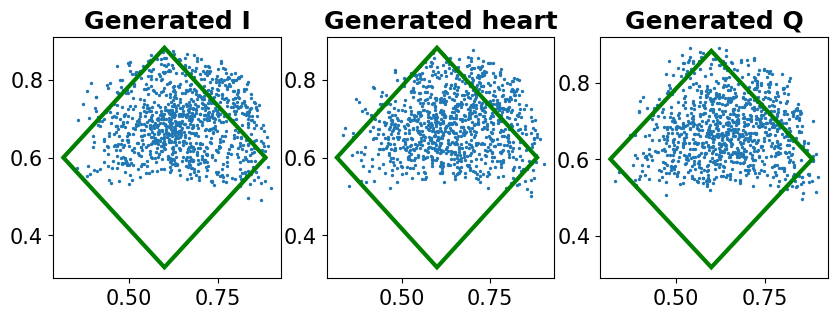}
         \label{fig:2d_custom_QAN}
     }\quad
     \subfloat[Distribution generated by InfoQGAN]{
         \includegraphics[width=0.48\textwidth]{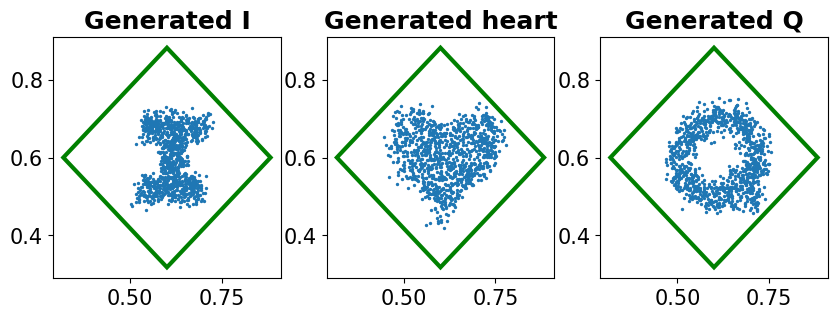}
         \label{fig:2d_custom_InfoQGAN}
     }
    \caption{\textbf{Comparison between the target distributions and those generated by QGAN and InfoQGAN.} This experiment demonstrates how feature disentanglement can be utilized. The same can be done with InfoGAN, but not with QGAN and GAN. Figure~\ref{fig:2d_custom_InfoQGAN_Q} shows an example where InfoQGAN mapped the target distribution to code space when creating the heart.}
    \label{fig:2d_custom_compare}
\end{figure}

The importance of the feature disentanglement ability lies in its versatility. In this case, the x and y coordinates of the points to be generated can be determined separately, allowing you to generate any arbitrary distribution you want within the diamond shape. To achieve this, you need to convert the target distribution into latent code space. First, translate $(0.6, 0.6)$ to $(0, 0)$. Next, normalize the magnitudes of both latent code vectors to 0.2. This process is illustrated in Figure~\ref{fig:2d_custom_InfoQGAN_Q}. Finally, perform a basis change so that they become $(0.5, 0)$ and $(0, 0.5)$ in the latent code space. The rest of the noise input can be taken from the $\mathcal{U}([-0.5, 0.5])^{3}$. We created the target distributions and let InfoQGAN generate it, and the results can be seen in the Figure~\ref{fig:2d_custom_compare}. As you can see, InfoQGAN could generate the arbitrary distributions we designed, but this was not possible in QGAN because the latent code has no meaning.

\begin{table}[t]
\centering
\begin{tabular}{|c|c|c|c|c|c|c|}
\hline
\rowcolor[HTML]{C0C0C0} 
Varient         & $D_{ks}$         & $p$-value          & Mutual Information & Angle             & $|v_{c_1}|$      & $|v_{c_2}|$      \\ \hline
GAN             & 0.10772          & 0.00081          & 0.73317            & 5.03206           & 0.51476          & 0.70147          \\ \hline
InfoGAN         & {\ul 0.05410}    & {\ul 0.27817}    & \textbf{3.34405}   & {\ul 88.04458}    & {\ul 0.98523}    & \textbf{0.99455} \\ \hline
QGAN~(ideal)     & 0.08609          & 0.01358          & 0.37553            & 15.49166          & 0.23942          & 0.56971          \\ \hline
QGAN~(noisy)     & 0.24800          & $1.65591\times 10^{-18}$      & 0.93158            & 11.32972          & 0.24764          & 0.51715          \\ \hline
InfoQGAN~(ideal) & \textbf{0.03843} & \textbf{0.70276} & \uline{2.98083}   & \textbf{89.40658} & \textbf{1.00538} & {\ul 0.95730}    \\ \hline
InfoQGAN~(noisy) & 0.22850          & $9.37677\times 10^{-16}$      & 2.39325            & 84.41651          & 0.99211          & 0.94513          \\ \hline
\end{tabular}
\caption{\label{tab:exp1result} \textbf{Comparative analysis of QGAN and InfoQGAN performance.} The angle metric represents the angle between $v_{c_1}$ and $v_{c_2}$. All metrics were computed based on the model from the epoch that achieved the highest $p$-value. In noisy environments, the $p$-value significantly decreased due to smaller, noise-distorted generated shapes. Evaluation of mutual information was performed by generating a sample set matching the size of the target distribution and re-training MINE to compute mutual information.} For each column, the best value is highlighted in bold, while the second-best value is underlined.
\end{table}


\subsection*{Iris dataset augmentation}
\begin{figure}[t!]
    \includegraphics[width=\textwidth]{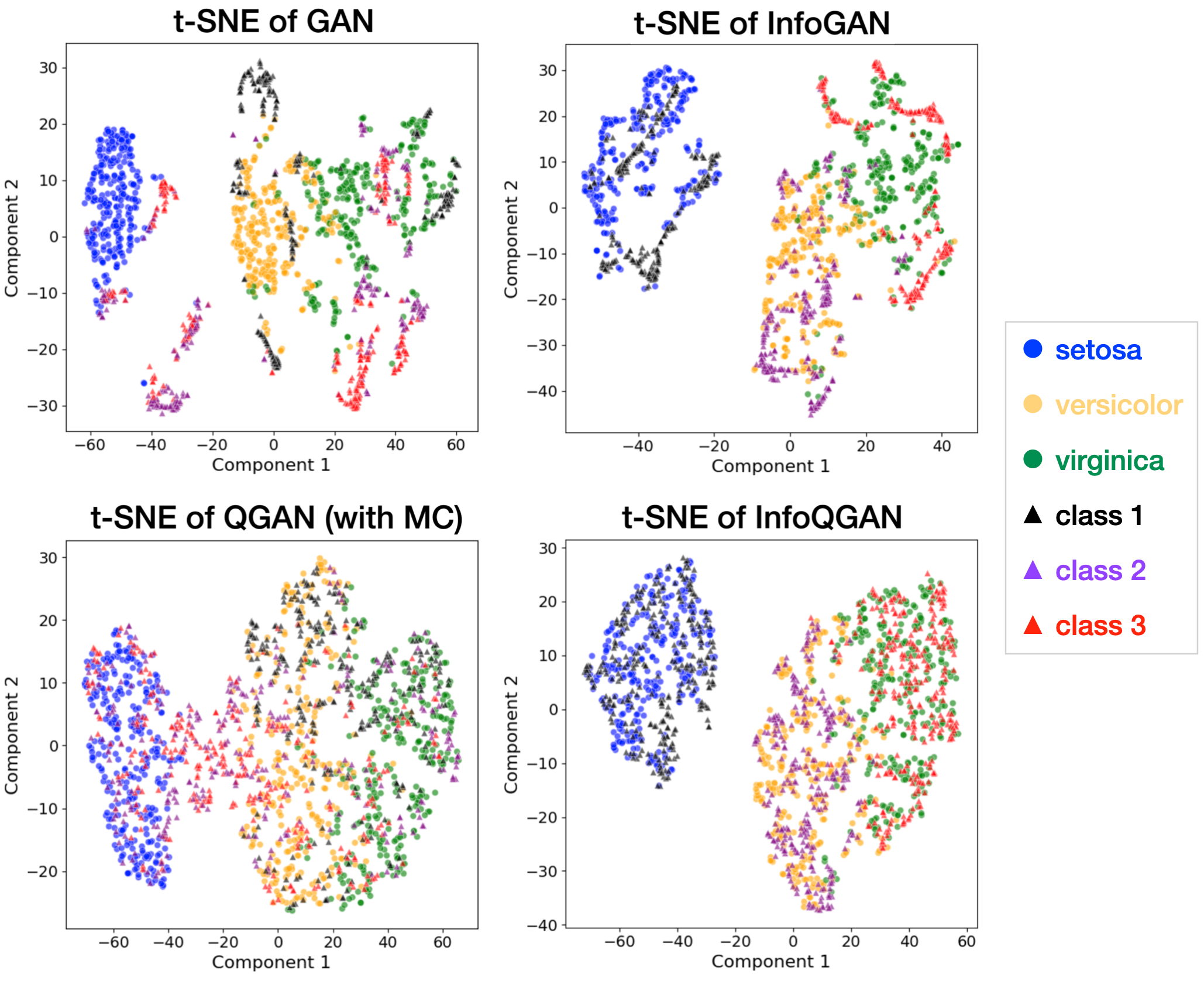}
    \caption{\textbf{Comparison of t-SNE visualizations.} The legend labels 1, 2, and 3 represent data generated by the generator, corresponding to $c_1$ values of -1.0, 0, and 1.0, respectively.}
    \label{fig:tsne_comparison}
\end{figure}

\begin{figure}[t!]
\includegraphics[width=\textwidth]{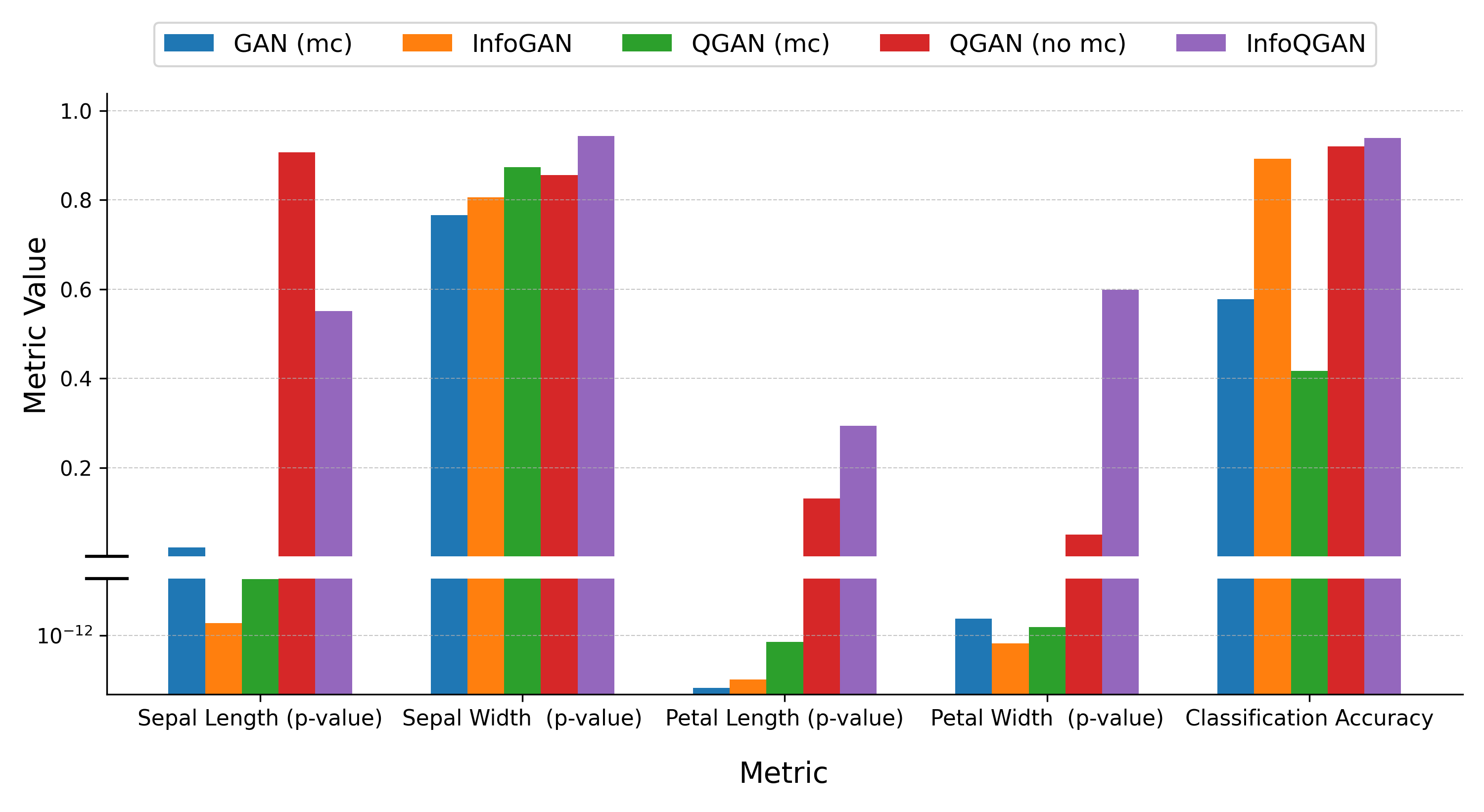}
\caption{\textbf{$p$-values and classification consistency across different models.}
The $p$-values for each attribute and classification consistency, calculated at the epoch with the highest total $p$-value.
Values below $10^{-3}$ are plotted on a logarithmic scale; see Table~\ref{sup-tab:decision_acc} for the exact numerical values.}
\label{fig:decision_acc}
\end{figure}

The IRIS dataset was first introduced in a seminal paper by R.A. Fisher~\cite{fisher1936use}. It contains measurements of the sepal length, sepal width, petal length, and petal width for three species of iris flowers: Setosa, Versicolor, and Virginica. Excluding the species label, the dataset consists of four numerical features, each with distinct ranges and distributions. Due to its low dimensionality, the IRIS dataset is widely used as a benchmark for classification models. However, the original dataset is limited to only 150 samples. To address this, we used the extended IRIS dataset~\cite{baladram2020iris}, which includes additional dimensions and comprises 1,200 samples. From this dataset, we retained only the sepal and petal length and width features, splitting 900 samples into the training set and 300 into the test set.  

In this experiment, we evaluated the ability of GAN, InfoGAN, QGAN, and InfoQGAN to generate synthetic samples resembling the IRIS training dataset. As in the previous experiment, the generator circuit structure for InfoQGAN and QGAN is identical, utilizing 5 qubits and 20 layers, amounting to 100 trainable parameters. The generator output is derived from the probability of each qubit being in the $|1\rangle$ state. However, since the distribution of each feature in the IRIS dataset varies, post-processing is required for the quantum models (unlike GAN and InfoGAN). Let $m_i$ and $M_i$ denote the minimum and maximum values of the $i$-th feature in the dataset, respectively, and let $p_i$ represent the probability of the corresponding qubit being in the $|1\rangle$ state. The transformed output is then defined as:
\begin{equation}
    p_i \rightarrow m_i + \frac{(p_i - 0.15)(M_i - m_i)}{0.7}.    
\end{equation}

The classical generator employs six hidden dimensions, resulting in a total of 106 trainable parameters. The input noise vector consists of a two-dimensional latent code and three additional noise dimensions. Since there are three iris species, the first latent code $c_1$ follows a categorical distribution:
\begin{equation}
    c_1 \sim \mathcal{U}(\{-1.0, 0, 1.0\}).    
\end{equation}

Thus, the generator’s seed vector $z$ is sampled as follows:  
\begin{equation}
    z \sim \mathcal{U}([-1.0, 1.0])^3 \otimes \mathcal{U}(\{-1.0, 0, 1.0\}) \otimes \mathcal{U}([-1.0, 1.0]).    
\end{equation}

For the classical generator, we expanded the input seed range by a factor of 50 to prevent instability during training caused by a narrow input range.  

For InfoQGAN and QGAN, the learning rates for the quantum generator and MINE were set to 0.003, while the discriminator was trained with a learning rate of 0.0003. For InfoGAN and GAN, the classical generator and MINE were trained with a learning rate of 0.001, and the discriminator used a learning rate of 0.0002. The training parameter $\beta$ was fixed at 0.04. A summary of the training hyperparameters is provided in Table ~\ref{sup-tab:iris-gen-hparams}, which is included in the supplementary material.

Figure~\ref{fig:tsne_comparison} presents a two-dimensional t-SNE visualization of the distributions generated by each model alongside the target distribution. The visualization indicates that InfoQGAN produces a distribution that closely aligns with the IRIS dataset. The results of the KS test for equality of distributions across each feature dimension are summarized in Figure~\ref{fig:decision_acc}. Three distinct types of IRIS are generated depending on the value of $ c_1 $, demonstrating their feature disentanglement capabilities. In contrast, both GAN and QGAN experience mode collapse in certain cases, rendering the relationship between $ c_1 $ and the IRIS type relatively ambiguous.

To quantify the consistency of the relationship between $ c_1 $ and the species of the generated data, we employed a decision tree model pre-trained on the training dataset. After classifying the generated data using this pre-trained decision tree, we assigned species labels to $ c_1 $ via majority voting and measured the classification accuracy. Figure~\ref{fig:decision_acc} presents the KS test results, including classification consistency. Among the compared models, InfoQGAN exhibits the best alignment with the target distribution.

Finally, we evaluated the performance of a classification model trained on augmented data. The evaluation process is as follows:
\begin{enumerate}
    \item Train a secondary classification model on the training dataset.
    \item Extract $ m $ (where $ m $ is a multiple of 3) generated samples, ensuring even distribution across different values of $ c_1 $, and classify them using the trained classification model.
    \item Determine the species corresponding to each $ c_1 $ value via majority vote and assign labels accordingly.
    \item Combine the labeled generated data with the training dataset to create an augmented dataset, train the main classification model, and evaluate its accuracy on the test dataset.
    \item Repeat steps 1 through 4 a total of 100 times and compute the average accuracy.
\end{enumerate}

\begin{figure}[t]
\centering
\subfloat[Decision Tree]{
\includegraphics[width=0.3282\textwidth]{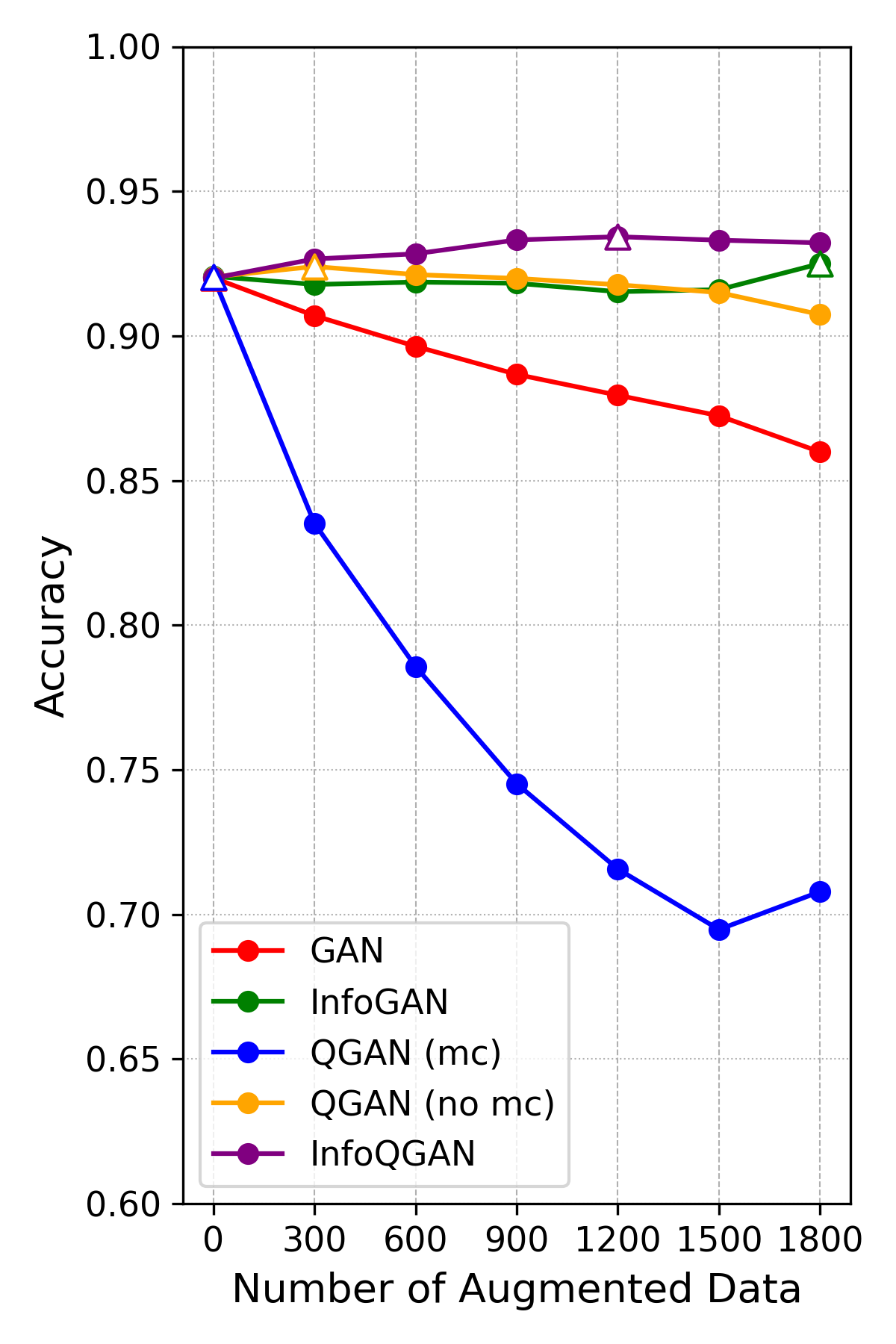}
\label{fig:acc_decision_tree}
}\quad
 \subfloat[Logistic Regression]{
\includegraphics[width=0.3008\textwidth]{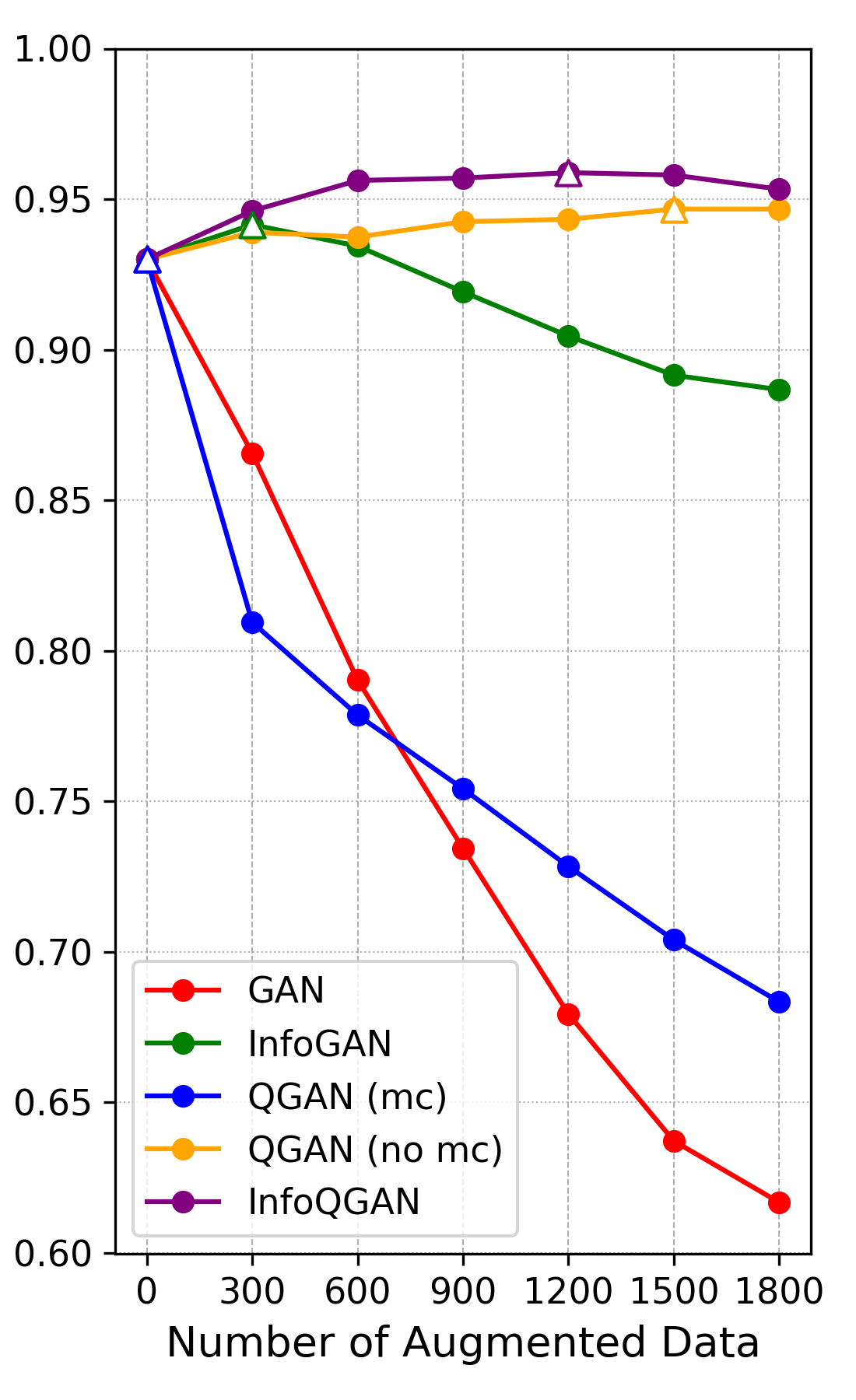}
\label{fig:acc_logistic_regression}
}\quad
\subfloat[k–NN]{
\includegraphics[width=0.3008\textwidth]{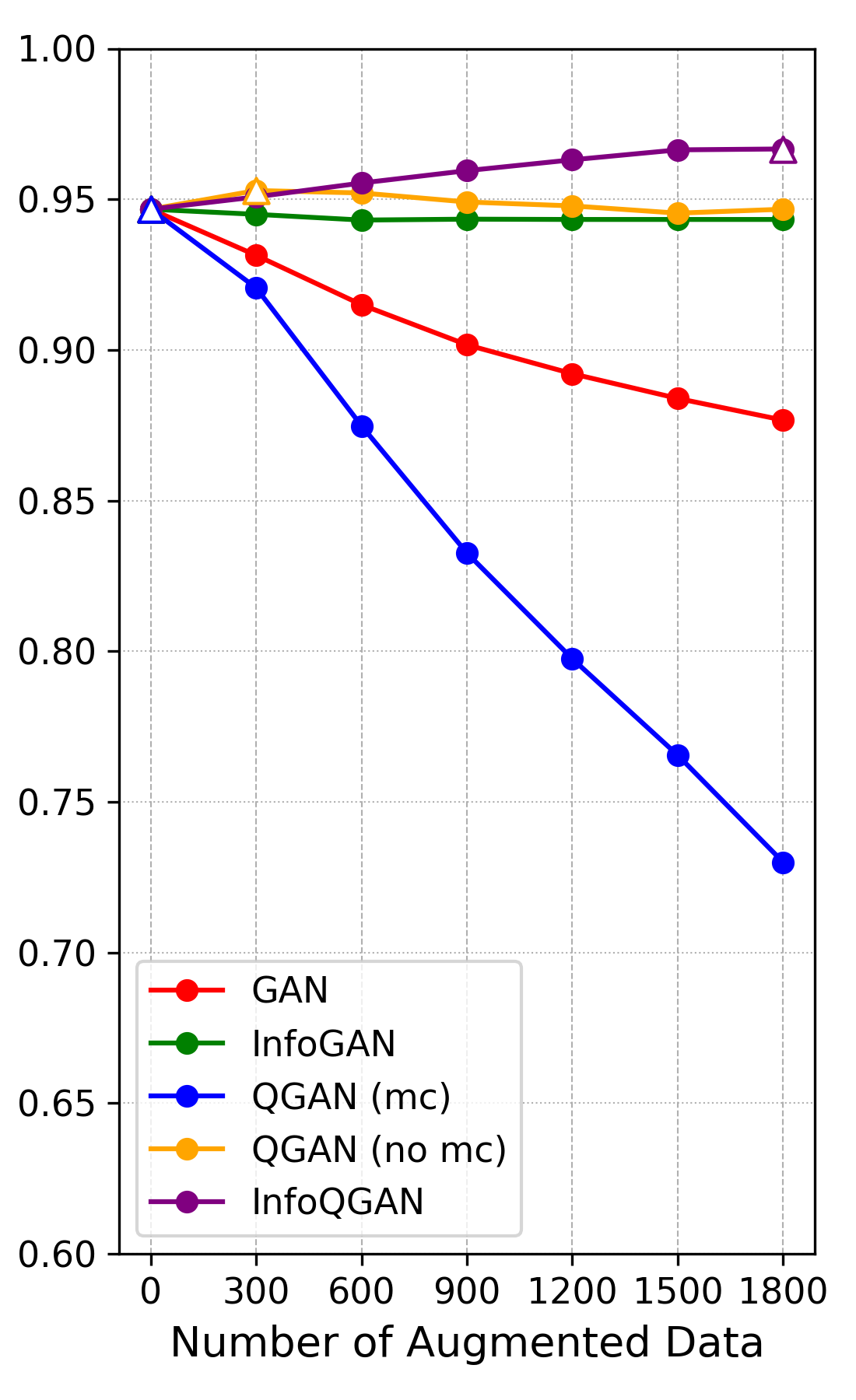}
\label{fig:acc_k_nn}
}
\caption{\textbf{Accuracy of classifiers trained on augmented data from different generative methods.} Each subplot corresponds to one of three classification algorithms (Decision Tree, Logistic Regression, k–NN) and shows its accuracy when trained on datasets augmented with samples from one of five generative models (GAN, InfoGAN, QGAN with mode collapse, QGAN without mode collapse, InfoQGAN). The horizontal axis indicates the number of augmented samples added, and white triangles denote each classifier’s peak accuracy. Exact numerical values are provided in Table~\ref{sup-tab:exp2result}, which is included in the supplementary material.}
\label{fig:exp2result}

\end{figure}

We evaluated performance using widely used classification models, including Decision Tree, k-Nearest Neighbors (KNN), and Logistic Regression. The numerical results are summarized in \textcolor{blue}{Figure}~\ref{fig:exp2result}. Here, InfoQGAN demonstrates the strongest data augmentation capability. Augmenting the dataset with InfoQGAN led to performance improvements of 1.43pp, 2.00pp, and 2.88pp across the three classification models. However, as with other data augmentation techniques, excessive augmentation relative to the original training data can lead to diminishing or even negative returns in performance.

The comparison between QGAN models with and without mode collapse reveals that mode collapse negatively impacts augmentation performance. This is because the distribution of the augmented data deviates from that of the original dataset. This effect is particularly evident in QGAN (with mode collapse) and GAN, where classification accuracy continuously declined as more augmented data was introduced. In contrast, InfoQGAN allowed for better alignment between the augmented and original distributions by generating samples with a balanced 1:1:1 ratio for different values of $ c_1 $, thereby achieving superior performance. This approach works because InfoQGAN does not cause mode collapse and allows for feature disentanglement.

\section*{Discussion}
Mode collapse is a critical issue in generative models, particularly in the generator component, as it directly impacts data augmentation. Our experiments highlight that appropriately setting the \(\beta\) value is crucial for effective training. If \(\beta\) is too large, the model focuses excessively on maximizing mutual information rather than generating the target distribution, which hampers learning. Furthermore, the convergence of mutual information during training did not exhibit a strict one-to-one proportional relationship with the \(\beta\) coefficient.

In the case of quantum generators, the range of the input seed significantly affected performance. Since unitary operations preserve the inner product of input states, generating distributions with high variance requires increasing the seed range, whereas generating distributions with lower variance requires reducing it. This highlights the importance of careful initialization when working with quantum circuits.

In the NISQ era, the quantum components of a system are less accurate and significantly slower than their classical counterparts. Furthermore, reducing the error rate requires executing more quantum circuits, scaling as \( O(1/\epsilon) \) shots. To mitigate these errors, it’s critical to minimize the use of two-qubit gates, and adopting trainable entanglement structures can be especially helpful. Alternatively, one can optimize performance by shifting additional work onto classical processors without increasing quantum operations. One promising approach is quantum error mitigation~\cite{endo2018practical, endo2021hybrid, cai2023quantum}. In this context, a key advantage of InfoQGAN is that it enhances performance without modifying the quantum part of QGAN. Additionally, the structural simplicity of the MINE model facilitates scalability while maintaining low runtime and computational complexity. However, in the Fault-Tolerant Quantum Computing (FTQC) era, the scalability of the classical MINE model will reach its limits. In such cases, quantum mutual information neural estimation (QMINE)~\cite{shin2023estimating, goldfeld2023quantum} can be employed to compute quantum mutual information between inputs and outputs, ensuring the retention of quantum advantages.

Notably, existing QGAN architectures have yet to achieve effective feature disentanglement. By introducing InfoQGAN, we overcome this limitation and enable more precise control over generated outputs. For instance, if InfoQGAN is applied to probability distribution function (pdf) modeling as in~\cite{zoufal2019quantum}, it could be used to generate weighted combinations of multiple distributions, which may find applications in financial modeling. If InfoQGAN is applied to quantum-data tasks, such as quantum state preparation or clustering Hamiltonian ground states to identify unknown quantum ground states. In this context, InfoQGAN could be employed to disentangle the basis of ground states, providing clearer structural insights. Furthermore, comparing InfoQGAN with other enhanced QGAN variants would be a valuable direction for future research, offering deeper insights into its relative strengths and practical applicability.

\section*{Data availability}
The data and software that support the findings of this study can be found in the following repository: \url{https://github.com/red1108/InfoQGAN}.

\section*{Acknowledgements}
This work was supported by the National Research Foundation of Korea (NRF) through grants funded by the Ministry of Science and ICT (NRF-2022M3H3A1098237; RS-2024-00404854; RS-2025-00515537). This work was partially supported by an Institute for Information \& Communications Technology Promotion (IITP) grant funded by the Korean government (MSIP) (No. 2019-0-00003; Research and Development of Core Technologies for Programming, Running, Implementing, and Validating of Fault-Tolerant Quantum Computing Systems), and Korea Institute of Science and Technology Information (KISTI).

\section*{Author contributions}
M.L. conducted the main analysis, including numerical simulations, performed data visualization, and prepared the initial draft of the manuscript. M.S. proposed the initial draft idea and verified the computations. J.L. analyzed the results and contributed to the theoretical framework. J.L. and K.J. supervised the overall project, provided conceptual guidance. All authors contributed to the writing and discussions.

\section*{Competing interests}
The author J.L. is employed by Norma Inc., but there is no conflict of interest related to this work. The remaining authors declare that the research was conducted in the absence of any commercial or financial relationships that could be construed as a potential conflict of interest.
\end{document}